\documentclass[12pt]{article}
\usepackage{amsfonts,amssymb}
\usepackage{graphicx,epsfig}
\newcommand{\ket}[1]{\left\vert #1  \right\rangle}
\newcommand{\average}[1]{\langle #1 \rangle}
\newcommand{\bra}[1]{\left\langle #1  \right\vert}

\newcommand{\Tr}{{\hbox{Tr\,}}}

\newcommand{\form}{{\rho}}
\begin{document}
\author{O. Gat and J.E. Avron}
\title{Semiclassical analysis and the magnetization of the Hofstadter
model}

\maketitle
\begin{abstract}
The magnetization and the de Haas-van Alphen oscillations of Bloch
electrons are calculated near commensurate magnetic fluxes. Two
phases that appear in the quantization of mixed systems---the
Berry's phase and a phase first discovered by Wilkinson---play a
key role in the theory.
\end{abstract}

The magnetization of a free electron gas was calculated by L.
Landau in 1930 in the early days of quantum mechanics
\cite{ref:ll}. Considerable efforts have since been devoted to
extending Landau results to Bloch electrons, {\it i.e.}, in the
presence of periodic background potential. Most of the efforts and
progress made was in the region of weak magnetic fields
\cite{ref:peierls,ref:blount} where the flux $\Phi$ through a unit
cell is small. This is adequate for most solid state applications.
There is, however, also considerable interest in a better
understanding of phenomena that have to do with commensuration in
condensed matter physics
\cite{ref:commensurate,ref:azbel,ref:last,ref:guarneri,ref:wilkinson}.
This is the case when $\Phi$ is close to a rational number. The
magnetization of Bloch electrons near rational fluxes, other than
$\Phi=0$, remained an open challenge which we solve here. The
difficulty lies in the delicate spectral properties resulting from
commensuration \cite{ref:azbel}.  The Hofstadter model is a basic
model for a system where commensuration plays a role. It is also a
basic paradigm for a quantum system with fractal spectra
\cite{ref:last}, anomalous quantum transport \cite{ref:guarneri}
and the (integer) quantum Hall effect \cite{ref:tknn}.

The problem of magnetization near fractional flux becomes
tractable by an idea that goes back to M.~Wilkinson
\cite{ref:wilkinson}. Namely, that near a rational flux  the
Hamiltonian can be understood as the semiclassical quantization of
a mixed system: In mixed systems some, but not all, degrees of
freedom may be treated semiclassically. As a consequence the
``classical Hamiltonian'' is matrix (or operator) valued. Pauli
and Dirac equations for a spinning electron in a slowly varying
potential, the Born-Oppenheimer theory of molecules, and the
Hofstadter model near rational flux \cite{ref:wilkinson} are
examples of mixed systems. In the Hofstadter model the role of the
Planck constant is played by the deviation from a nearby rational
\begin{equation}\label{eq:h-def}h=\Phi-\frac p q\ .\end{equation}

Littlejohn and Flynn \cite{ref:littlejohn} developed an elegant
geometric formalism for the quantization of mixed systems. They
show that in order $\hbar$ the quantization of mixed systems gives
rise to two phases: One is the Berry's phase \cite{ref:berry} and
the other is a phase that is sometimes known as the ``no-name
phase'' \cite{ref:weinstein} and sometimes as the Wilkinson-Rammal (WR)
phase~\cite{ref:spohn}. For both phases to appear, the ``classical
Hamiltonian'' must have non-trivial commutation properties in both
coordinates and momenta. These phases play a central role in
determining the magnetization, see e.g. Eq.~(\ref{eq:mc-ab})
below.

Our results are closely related to recent progress made in the
semiclassical dynamics of Bloch electrons under slowly varying
electric and magnetic fields \cite{ref:spohn}. When one goes
beyond the leading order expressed by Peierls substitution
\cite{ref:peierls,ref:spohn} one finds that the Berry's phase and
the WR phase play a role in the dynamics. This lead
\cite{ref:spohn} to identify the WR term with the magnetization of
a wave packet. Although related, the notions of wave packet vs. 
thermodynamic magnetization, expressed in Eq.~(\ref{eq:mc-ab}), 
are distinct; for example, wave packet magnetization is not 
defined in the gaps, while the thermodynamic magnetization and, 
of course, the de Haas van Alphen oscillations, have a 
non-trivial dependence on the chemical potential also in the gaps.

\begin{figure}[ht]
\includegraphics[width=13.cm]{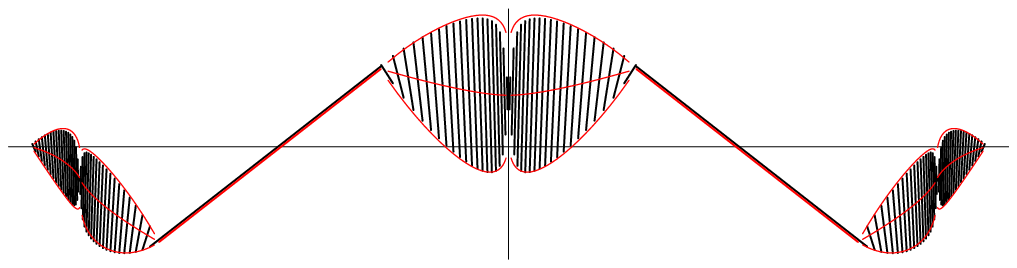}\vskip3mm
\hbox{\includegraphics[width=5.5cm]{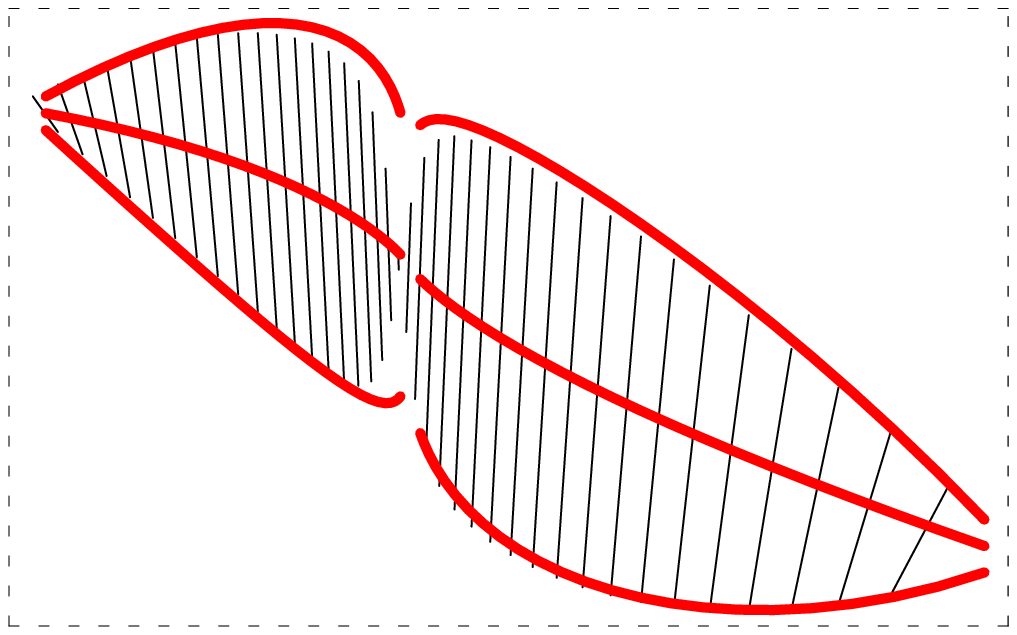}\hskip5mm
\includegraphics[width=5.5cm]{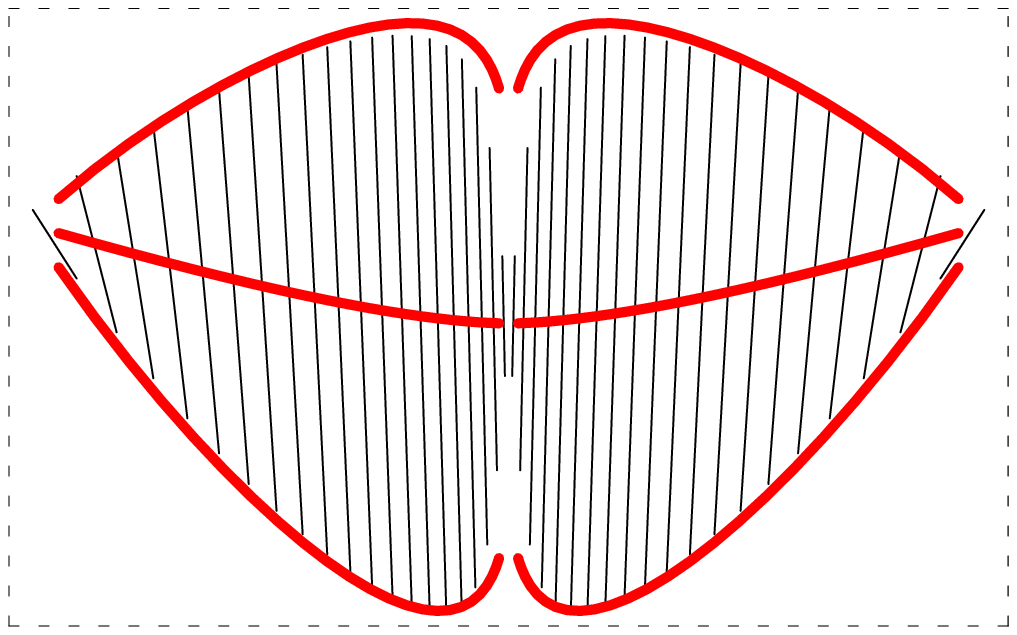}}
\caption{The de Haas van-Alphen oscialltions of the magnetization
as a function of the chemical potential in the Hofstadter model for
flux $\Phi=40/121$, compared with the limiting envelopes and their averages, and the gap magnetization for $\Phi\to1/3$. The two insets are details of the left and center envelopes.} \label{fig:40/121}
\end{figure}

Fig.~\ref{fig:40/121} shows the zero temperature magnetization at 
$\Phi=\frac 1{3+\frac 1 {40}}$. The complexity of the 
magnetization is due to the multiplicity of scales: The big scale 
is determined by the denominator $q=3$, and the  small scale by 
$h$. On the small scale one sees the rapid de Haas-van Alphen 
oscillations. On the big scales one sees continuous features: the 
linear pieces in the (big) gaps, the envelopes of the de Haas-van 
Alphen oscillations and their mean. Our theory of magnetization 
accounts for all these features.

We show that while the amplitude of the oscillation is determined
solely by the leading terms in the semiclassical expansion, the
mean magnetization requires knowledge of the terms beyond leading
order, thus depending on the fine details of the spectrum.
Intriguingly, it is the {\em latter} quantity which is stable
against perturbations. Finite temperatures larger than the typical
eigenvalue spacing wash out the de Haas-van Alphen oscillations of
the small scale but leave intact the mean magnetization.
Semiclassical approximations that retain only the leading order
yield {no} magnetization at all at finite temperatures.

Let us start by describing the semiclassical quantization of
mixed systems~\cite{ref:littlejohn}. The ``classical
Hamiltonian'', ${\cal H}(x,k)$, is a Hermitian {\em matrix} which
depends on $x$ and $k$.  We shall denote by $\varepsilon_j(x,k)$
its $j$-th band of eigenvalues and by $\ket{u_j(x,k)}$ the
corresponding eigenvectors. We shall also assume that bands do not
cross\footnote{In the Hofstadter model, this is guaranteed by
Chambers relation.}. The corresponding quantum Hamiltonian is
${\cal H}(\hat x,\hat k)$ with $[\hat x,\hat k]=i\hbar$.

Let $S(E;H,\form)$ denote the classical action associated with a
closed orbit of energy $E$ of a classical Hamiltonian  function
$H$, with phase space area form $\form$. Since phase space is two
dimensional the action is the area enclosed by the orbit. Note 
that $H$, unlike ${\cal H}$, is a scalar valued function. The 
Bohr-Sommerfeld quantization rule in mixed systems says that the 
semiclassical approximation of the eigenvalue $E_n$ is given by
\begin{equation}\label{eq:bs}
S(E_n;H_\hbar,\form_\hbar)=h\,(n+\gamma_m)\ ,\quad n\in \mathbb{Z}
\end{equation}
where $\gamma_m$, is the Maslov index of the
orbit~\cite{ref:maslov}. $H_\hbar$ has an expansion in powers of
$\hbar$, $H_\hbar=H^{(0)}+\hbar H^{(1)}+\cdots\,$. Peierls
substitution sets $H_j^{(0)}(x,k)=\varepsilon_j(x,k)$, and the
next order is~\cite{ref:littlejohn}
\begin{equation}\label{eq:bandbs}
H_j^{(1)}(x,k)= {\rm Im} \bra{\partial_xu_j}({\cal
H}-\varepsilon_j)\ket{\,\partial_ku_j}(x,k)\ .
\end{equation}
The expansion $\form_\hbar=\form^{(0)}+\hbar \form^{(1)}+\cdots$
begins with the canonical form $\form_j^{(0)}=dk\wedge dx$ and the
subleading term is the Berry curvature form
\begin{equation}\label{eq:phase-space}
\form_j^{(1)}=2\omega(x,k)\, dk\wedge dx\ ,\quad \omega(x,k)={\rm
Im}\bra{\partial_xu_j}\partial_k{u_j}\rangle(x,k) \
.\end{equation}

This formulation is manifestly gauge invariant (independent of the
choice of phases for $\ket{u_j}$) and preserves the symmetry
properties of $\cal H$, which is useful when one wants to
correctly count the dimension of the Hilbert space of the
quantized operator, as we now proceed to explain.

Suppose that $\cal H$ is periodic in both $x$ and $k$ up to gauge
transformations, and hence describes (classical) motion on a phase
space torus $\cal T$. $H_j^{(0)}$, $H_j^{(1)}$ and $\form_j^{(1)}$
are all well-defined functions on $\cal T$. The Chern number of
the $j$-th band is the integer
\begin{equation}\label{eq:chern}
C_j=\frac 1 {\pi}\int_{\cal T}\,\omega_j  \, dk\wedge dx
\end{equation}
 It follows from
Eqs.~(\ref{eq:bs},\ref{eq:phase-space}) that the dimension of the
Hilbert space associated with the $j$-th band is
\begin{equation}\label{eq:dimensions}
 \frac{|{\cal T}|}h+C_j\ ,
 \end{equation}
where $|{\cal T}|$ denotes the area of $\cal T$. Since the
dimension of the Hilbert space is necessarily an integer,
quantization on the torus is possible only for certain values of
$h$. Eq.~(\ref{eq:dimensions})  goes beyond the classical Weyl law
which only determines the leading, $|{\cal T}|/h$, behavior. The
Chern numbers shift states between the spaces of different bands
since $\sum C_j=0$ \cite{ref:ass}.

The $\hbar$ corrections to the action can be moved from the
left hand side of the Bohr-Sommerfeld relation to the right hand
side, where they acquire an interpretation as two additional phase:
\begin{equation}\label{eq:bss1}
S_j(E)= h\,\big( n+\gamma_m+\Gamma_j(E)\big),\quad
-2\pi\Gamma_j(E)= \gamma_j^b(E)+ \gamma_j^{wr}(E),\end{equation}
where $S_j(E)=S(E;\varepsilon_j,dk\wedge dx)$. $\gamma_j^b(E)$ is
the Berry's phase \cite{ref:berry},
\begin{equation}\label{eq:berry-phase}\gamma_j^b(E)=2
\,\int\, \omega_j\,\theta(E-\varepsilon_j)\,dk\wedge dx,
\end{equation}
 and $\gamma_j^{wr}(E)$ is the Wilkinson-Rammal phase
\begin{equation}\label{eq:flw-phase}
\gamma_j^{wr}(E)=-\int H^{(1)}_j\,
\delta\big(E-\varepsilon_j\big)\,dk\wedge dx.
\end{equation}
It is noteworthy that the WR phase {\em need not} vanish at band
edges.

Let us now recall some basic facts about the Hofstadter
model~\cite{ref:azbel,ref:wilkinson}. When the magnetic flux
$\Phi$ is a rational number $p/q$, the model is represented by
\cite{ref:wilkinson}
\begin{equation}\label{eq:hofstadter} {\cal H}(x,k)=e^{2\pi i x}
U+e^{2\pi ik} V+{\mathrm h.c.}
\end{equation}
where $U$ and $V$ are the $q\times q$ matrices
\begin{equation}\label{eq:pqrep}
U=\pmatrix{0&1&\cdots&0\cr&\ddots&\ddots&\cr&&\ddots&1\cr1&&&0}\ ,
\quad V=\pmatrix{1&&&\cr&e^{2\pi ip/q}&&\cr&&\ddots&\cr&&&e^{2\pi
i(q-1)p/q}\cr} \ .\end{equation}
The magnetic bands of the Hofstadter Hamiltonian at $\Phi=p/q$ are given by
$\varepsilon_j(x,k)$ on the Brillouin zone
\begin{equation}\label{eq:reduceddzone}
BZ=\{x,k|\, 0\le x\le 1, \ 0\le k \le 1/q\}.
\end{equation}

Evidently, ${\cal H}(x,k)$ is periodic with period 1 in both
variables. Moreover, ${\cal H}$ is periodic with smaller periods
up to unitary transformations:
\begin{equation}\label{eq:periodicity}
 {\cal H}(x,k)=T  {\cal H}(x+1/q,k)T^\dagger=
G {\cal H}(x,k+1/q)G^\dagger
\end{equation}
$G$ is a gauge transformation (a diagonal unitary) and $T$ a
shift. This makes the band dispersion functions
$\varepsilon_j(x,k)$ periodic with periods $1/q$ in each variable
and with $q$ periods in ${\cal T}= BZ$.

The spectrum of the Hofstadter model for other values of $\Phi$ is
obtained  by setting $[\hat x,\,\hat k]=i\hbar$  in $\cal H$ with
$\hbar$ given by Eq.~(\ref{eq:h-def}). $BZ$ is the minimal torus
on which $\cal H$ may be quantized.  The dimension of the Hilbert
space associated with it is then $\frac 1{qh}-C_j$. Since the band
function are $q$-periodic on $BZ$, the number of {\em distinct}
eigenvalues is of order $1/(q^2h)$. The semiclassical
approximation is valid provided this number is large i.e.
$q^2\hbar\ll 1$.

We now turn to the magnetization of the model. Recall that the
Hofstadter model approximates the Schr\"odinger equation in two
dual limits: When the magnetic field is weak relative to the
periodic potential and also in the opposite limit where the magnetic
field dominates all other interactions. The two limits have
related but different thermodynamics. For the sake of concreteness
we shall consider the tight-binding interpretation. The
magnetization of the ``split Landau level'' follows from the
duality transformation of~\cite{ref:gat}.

The thermodynamic potential per lattice site of the Hofstadter
model for rational flux and zero temperature is \cite{ref:gat}
\begin{equation}\label{eq:omega} \Omega(\mu,\Phi)=\int_{BZ} dk\wedge dx\,\hbox{\Tr}
\bigg(\mu-{\cal H}\bigg)_+
\end{equation}
where $x_+=x\theta(x)$. When $\mu$ is in a spectral gap, the
thermodynamic potential can be written as a sum of the potential
of the occupied bands, $\Omega=\sum_< \Omega_<$.

The  magnetization per unit area $m$ is\footnote{ To translate the
magnetization to ordinary units one needs to divide our
dimensionless magnetization by the unit of quantum flux.}
\begin{equation}\label{eq:rh-m-s}
m(\mu,\Phi)=- \left(\frac{\partial \Omega}{\partial
\Phi}\right)_\mu.
\end{equation}
The magnetization in the gaps can likewise be expressed as a sum
of the magnetization of the occupied  bands
\begin{equation}\label{eq:sum-mag} m(\mu,\Phi)=\sum_< m_<(\Phi)\ ,\end{equation}
where the magnetization of a full band is, as we
shall see below, (see also~\cite{ref:gat}):
\begin{equation}\label{eq:band-mag} m_j(\mu,\Phi)=-\frac 1 {2\pi }
\int_{BZ}\Big(2(\mu-\varepsilon_j)\omega_j-H^{(1)}\Big)\, dk\wedge
dx.\end{equation} The term proportional to $\mu$ is the Chern
number of the band. It follows that the magnetization as a
function of $\mu$ has quantized slopes in the gaps.

The envelope of the de Haas-van Alphen oscillations of $j$-th
band, as we shall show below, is given by
\begin{equation}\label{eq:lip}
 L_j(\mu,\Phi)=\average{m}_j(\mu,\Phi)\pm \delta m_j(\mu,\Phi).
\end{equation}
$\average{m}_j(\mu,\Phi)$, is the natural restriction of
Eq.~(\ref{eq:band-mag}) to a partially filled band, {\it i.e.},
\begin{equation}\label{eq:average-mag}
\average{ m}_j(\mu,\Phi)=-\frac 1 {2\pi}
\int_{BZ}\theta(\mu-\varepsilon_j)\Big(2(\mu-\varepsilon_j)\omega_j-H_j^{(1)}\Big)\,
dk\wedge dx.\end{equation} It describes the mean value of the
magnetization, averaged over the de Haas-van Alphen oscillations.

The width of the envelope is given in terms of the classical action
associated with $\varepsilon_j$:
\begin{equation}\label{eq:me}
\delta m_j(\mu )=\frac q 2\, {S_j(\mu)\over S_j'(\mu)}\ .
\end{equation}
$\frac 1 q \,S'(\mu)$ is proportional to the density of states.
Since the density of states in two dimensions diverges
logarithmically near the separatrix, the width $\delta m$ shrinks
to zero logarithmically there. Near the bottom of the band $S_j$
vanishes linearly while the density of states approaches a
positive value. This shows that $\delta m$ vanishes linearly at
band edges. These  properties characterize the universal lip-like
shape of the envelopes.

We conclude with an outline of the derivation of
Eqs.~(\ref{eq:lip},\ref{eq:average-mag},\ref{eq:me}). Consider the
zero temperature thermodynamic potential associated with one fixed
band $j$. It follows from the Chambers relation~\cite{ref:chambers} and the
square symmetry of the Hofstadter Hamiltonian that for all
energies except the separatrix, the level sets of
$\varepsilon_j(x,k)$ are deformed circles, and therefore $\gamma_m=1/2$.
All spectral quantities below refer to the same band,
and we may therefore suppress the index $j$ without risk of
confusion.

Suppose that $\mu$ is such that $n$ spectral points of the split $j$-th
band are occupied. Recall that each spectral point is $q$-fold
degenerate and that, by Eq.~(\ref{eq:dimensions}), $nq\le\frac 1
{qh}+C$. Suppose for definiteness that $\mu$ is below the
seperatrix. By the Bohr-Sommerfeld rule the thermodynamic
potential is (to order $\hbar^2$)
\begin{equation}\Omega(\Phi,\mu)=qh\,\left(\mu n-
\sum_{\ell=0}^{n-1}S^{-1}\bigg(\big(\ell+1/2+\Gamma(E_\ell)\big)
h\bigg)\right),\end{equation} The overall factor $qh$ comes from
the degeneracy per unit area of each eigenvalue. Approximating the
sum with the second Euler-Maclaurin sum formula gives (again to
order $\hbar^2$)
\begin{equation}\Omega(\Phi,\mu)=q\,\mu nh-q\int_0^{nh}dx\,
S^{-1}(x+h\Gamma(S^{-1}(x))) \ .\end{equation} Taking derivative
with respect to $\Phi$ is the same as taking derivative with
respect to $h$. The magnetization is therefore given (to order
$\hbar$) by
\begin{eqnarray}\label{eq:mj}\nonumber
m(\Phi,\mu)&=&-qn\left(\mu -S^{-1}\big((n+\Gamma)h\big)\right)+
q\,\int^{\mu}\Gamma(E)dE\\
&=&q\,n\left({E_n+E_{n-1}\over2}-\mu\right)+
q\,\int^\mu\Gamma(E)dE\ ,\end{eqnarray}
where the $E$-integration is over the $j$-th band energies below $\mu$.

The first term describes the de Haas-van Alphen oscillations:  It
vanishes in the middle of each spectral gap $[E_{n-1},E_n]$, and
reaches its maximum magnitude at the band edges. $\delta m(\mu)$ is half the variation of $m$ across a spectral gap.
Therefore
\begin{equation}\label{eq:delta-m}
\delta m(\Phi,\mu)= qn\frac{E_n-E_{n-1}}2\mathop{\longrightarrow}_{h\to0}
\frac{q\, S(\mu)}{2S'(\mu)}
\end{equation}

The first term in Eq.~(\ref{eq:mj}) has zero average over the gap,
while the second term is nearly constant. The mean magnetization
is therefore
\begin{equation}
\label{eq:mc-ab}\average{m}(\Phi,\mu)= -\frac q {2\pi}
\int^\mu\bigg(\gamma^b(E)+\gamma^{wr}(E)\bigg)\,dE
\end{equation}
The Berry's phase contributes
\begin{eqnarray}
q\int^\mu\gamma^b(E)\, dE&=&2\int_{BZ}\omega\, dk\wedge dx
\int^\mu\theta(E-\varepsilon) dE\nonumber \\
&=&2\int_{BZ}\,(\mu-\varepsilon)\omega\theta(\mu-\varepsilon)\,
dk\wedge dx\
\end{eqnarray}
The WR-phase contributes
\begin{eqnarray}
q\int^\mu\gamma^{wr}(E)\, dE&=&-\,\int_{BZ}dk\wedge dx
H^{(1)}\,\int^\mu
dE\,\delta(E-\varepsilon)\nonumber \\
&=&-\,\int_{BZ}H^{(1)}\,\theta(E-\varepsilon)\,dk\wedge dx
\end{eqnarray}
Together, they add up to give Eq.~(\ref{eq:average-mag}).

Finally, let us present a streamlined derivation of the rules for
band splitting \cite{ref:wilkinson}. Consider for example,
\begin{equation}
\Phi={1\over q-{1\over n}}={n\over qn-1}=
{1\over q}+h, \quad h={1\over q(qn-1)}\ ,
\end{equation} 
with $q$ odd and $n$ even. We demonstrate the following splitting rule: Of the $q$ bands associated with the flux $1/q$, the center band splits into $n-1$ subbands and the rest into $n$ subbands, together accounting
for the $qn-1$ band associated with the flux $\Phi$.
Clearly, this should follow from the dimension formula Eq.~(\ref{eq:dimensions}),
which requires the additional input of the Chern numbers.
The Diophantine equation of~\cite{ref:tknn} at flux $1/q$ bands 
implies that the Chern number of the center band is $1-q$, and all other
bands have Chern number $1$. Recalling that the area of the Brillouin zone
is $1/q$, Eq.~(\ref{eq:dimensions}) gives that the center band splits into
$q(n-1)$ levels and the other bands split into $qn$ levels each. We recall also that the band dispersion functions have $q$ periods in the Brillouin zone, and therefore there are only
$qn-1$ distinct levels, each $q$-fold degenerate\footnote{The levels
are broadened
into bands by tunnelling, which we do not discuss. This does not
modify the counting of dimensions.}.  This example illustrates the
algorithm which generates the hierarchical structure of the
Hofstadter butterfly.

{\bf Acknowledgment:} This work is supported by the Technion fund
for promotion of research and by the EU grant HPRN-CT-2002-00277.

\end{document}